# Individual Cell Fault Detection for Parallel-Connected Battery Cells Based on the Statistical Model and Analysis

Ziyou Song, Fanny Pinto Delgado, Jun Hou, Heath Hofmann, and Jing Sun

*Abstract*— Fault diagnosis is extremely important to the safe operation of Lithium-ion batteries. To avoid severe safety issues (e.g., thermal runaway), initial faults should be timely detected and resolved. In this paper, we consider parallel-connected battery cells with only one voltage and one current sensor. The lack of independent current sensors makes it difficult to detect individual cell degradation. To this end, based on the high-frequency response of the battery, a simplified fault detection-oriented model is derived and validated by a physics-informed battery model. The resistance of the battery string, which is significantly influenced by the faulty cell, is estimated and used as the health indicator. The statistical resistance distribution of battery strings is first analyzed considering the distribution of fresh and aged cells. A fault diagnosis algorithm is proposed and the thresholds (i.e., 2 standard deviation interval) are obtained through statistical analysis. Monte Carlo simulation results show that the proposed fault diagnosis algorithm can balance false alarms and missed detections well. In addition, it is verified that the proposed algorithm is robust to the uniform parameter changes of individual battery cells.

## I. Introduction

With the fast development of electrified vehicles and renewable energy systems, the lithium-ion battery has been widely used and intensively investigated [1]. Safety of Lithium-ion battery remains a critical concern, and is increasingly vital due to the continuous and significant improvement in battery energy density [2]. Fault diagnosis algorithms, especially ones which can timely detect initial faults to prevent severe damage, are extremely important to the safe and reliable operation of lithium-ion batteries [3]. The typical faults of the battery pack include, but are not limited to, significant degradation [4], internal/external short circuit [5], overdischarge/overcharge [6], and thermal runaway [7]. Compared to other fault types, the significant degradation of one individual cell has less influence on the battery string. It is essential to detect this kind of fault before it precipitates a severe failure.

Most studies on battery fault diagnosis mainly focus on series-connected battery cells. For example, Kong et al. [8] proposed a quantitative diagnosis method for the micro-short circuit fault of batteries. Gao et al. [9] used the mean-difference model to deal with the same fault. Ma et al. [10] conducted statistical analysis to determine abnormal voltages, and used the voltage as the health indicator for fault diagnosis. Note that the fault diagnosis for the series-connected battery string can be conducted based on the cell-level methods, since both the voltage and the current measurements of a cell are available.

In comparison, fault diagnosis for parallel-connected battery cells is difficult because generally only one voltage sensor and one current sensor are used for a string of parallel cells. The lack of current sensors for individual cells causes low observability, and therefore makes it difficult to detect an individual cell fault. Bruen and Marco [11] evaluated the imbalance of parallel-connected cells, and their experimental results showed a 30% difference in impedance results, a difference of 60% in peak cell current, and a difference of over 6% in charge throughput during cycling. Zhang et al. [12] first pointed out that the resistance of parallel-connected battery cells can be used as the health indicator for fault diagnosis. However, cell-to-cell variations were not considered hence the resistance thresholds, which are important to fault diagnosis, cannot be accurately quantified.

In this paper, a fault diagnosis algorithm for parallel-connected battery cells with one faulty cell, which has significant degradation when compared to the other cells, is proposed. The fault may be caused by manufacturing inconsistency, or uneven temperature distribution, which significantly influences battery degradation, or other factors. The resistance of the parallel battery string is selected as the health indicator, as the resistance is highly associated with battery degradation [13]. The statistical resistance distributions of both fresh and aged cells, which were obtained by characterizing 484 new and 1908 aged lithium-ion cells [14], are used to establish the statistical model for the battery string. Based on the statistical analysis, the thresholds in the fault diagnosis algorithm are carefully designed to balance false alarms and missed detections.

The rest of the paper is organized as follows. In Section II, the resistance of the battery strings is analyzed. In Section III, the resistance distribution of the battery string is presented and the thresholds for fault diagnosis are quantified. Conclusions are given in Section IV.

## II. Resistance of Battery Strings

Consider a battery string which consists of $N$ cells connected in parallel with one significantly degraded cell, as shown in Fig. 1. The voltages of all cells are equal and measured by the voltage sensor, assuming that the contact resistance is consistent and negligible. However, only the total current of the battery string is measured by the current sensor, and the current information for the individual cells is unknown.

*Research supported by U.S. Office of Naval Research (ONR) under Grants N00014-16-1-3108 and N00014-18-2330.

Z. Song and J. Sun are with the Department of Naval Architecture and Marine Engineering, University of Michigan, Ann Arbor, MI 48109, USA (e-mail: ziyou.songthu@gmail.com, jingsun@umich.edu).

F. Delgado, J. Hou, and H. Hofmann are with the Department of Electrical Engineering and Computer Science, University of Michigan, Ann Arbor, MI 48109 USA (e-mail: fapd, junhou, hofmann@umich.edu).

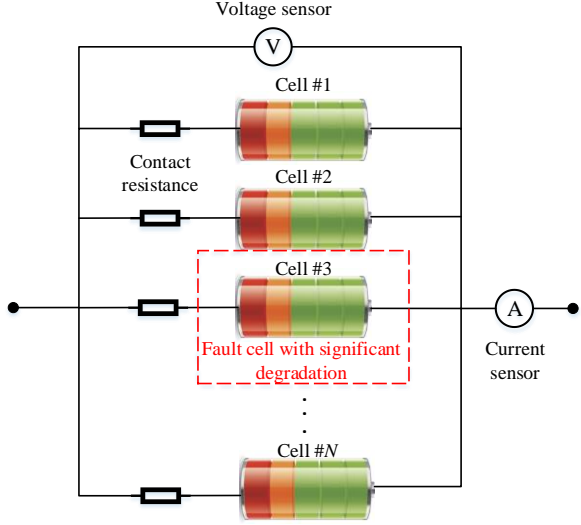

Figure 1. Parallel-connected battery cells with one faulty cell.

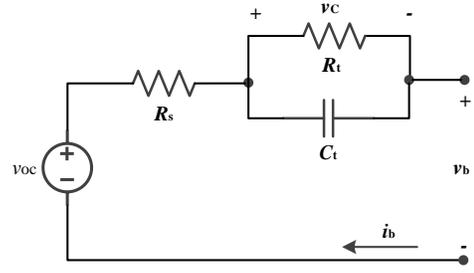

Figure 2. First-order equivalent circuit model for battery cell.

The ohmic resistance of a battery cell significantly increases with degradation. For example, the ohmic resistance of a Lithium-ion battery cell can increase by more than 200% when the cell capacity degrades to 85% of the initial value at 25°C [15]. We first investigate how the string resistance and other responses are impacted by the faulty cell. The first-order equivalent circuit model (ECM), as shown in Fig. 2, is adopted in the analysis. Note that the first-order ECM has been widely used due to its simplicity and sufficient accuracy [16]. The ECM dynamics can be described as:

$$\begin{cases} \dot{v}_C = -\frac{1}{\tau}v_C + \frac{R_t}{\tau}i_b, \\ v_b = v_{OC} - R_s i_b - v_C, \end{cases} \quad (1)$$

where $v_{OC}$ is the open circuit voltage (OCV), $v_C$ is the voltage of the RC pair, $v_b$ is the terminal voltage, $i_b$ is the battery current (positive for discharge and negative for charge), $R_s$ is the ohmic resistance, $R_t$ is the diffusion resistance, and $\tau$ is the time constant of RC pair. The OCV is determined by the battery SoC, and the SoC dynamic can be represented as:

$$z = z_0 - \int_0^t \frac{\eta}{3600 Q_b} i_b(t)\,dt, \quad (2)$$

where $z$ is the SoC, $z_0$ is the initial SoC, $\eta$ is the discharging/charging efficiency, $Q_b$ is the battery capacity, and $t$ is time (s). A linearized OCV-SoC relationship is used to simplify the analysis [17], which is described as:

$$v_{OC} = az + b, \quad (3)$$

where $a$ and $b$ are the constant coefficients. The relationship between $i_b$ and $v_b$ can therefore be given as [18]:

$$v_b(s) = \left[\frac{az_0}{s} + \frac{b}{s}\right] - \left[\frac{a}{s}\frac{\eta}{Q_b}i_b(s)\right] - \left[R_s i_b(s)\right] - \left[\frac{R_t}{1+\tau s}i_b(s)\right], \quad (4)$$

where $s$ is the complex Laplace variable. This shows that the terminal voltage dynamic includes four components associated with the initial SoC, the SoC variation, the ohmic resistance, and the RC pair, respectively. For lithium-ion batteries, it has been verified that these four components have significantly different responses in the frequency domain [19]. When the current frequency is relatively high (e.g., > 500mHz), the battery terminal voltage is dominated by the ohmic resistance. As a result, as long as the current profile contains enough high-frequency components, we can apply a high-pass filter to both the battery terminal voltage and the battery current and get the following equation.

$$\tilde{v}_b(s) = -R_s \tilde{i}_b(s), \quad (5)$$

where $\tilde{v}_b$ and $\tilde{i}_b$ are the filtered battery voltage and current. Based on the ECM for the battery cell, the model of the parallel connected cells is investigated. All cells in the battery strings have similar dynamics, as indicated in Eq. (4). Therefore, by incorporating the high-pass filter, the filtered voltage and the total current of the battery string (denoted as $\tilde{v}_{bs}$ and $\tilde{i}_{bs}$), which only consists of two cells, have the following relationship:

$$\tilde{v}_{bs}(s) = -\left(\frac{1}{R_{s1}} + \frac{1}{R_{s2}}\right)^{-1} \tilde{i}_{bs}(s), \quad (6)$$

where $R_{s1}$ and $R_{s2}$ are the ohmic resistance corresponding to two cells. Furthermore, the relationship shown in Eq. (6) can be generalized to a battery string including $N$ parallel cells, which can be given as:

$$\tilde{v}_{bs}(s) = -\frac{1}{\sum_{i=1}^{N}\frac{1}{R_i}} \tilde{i}_{bs}(s). \quad (7)$$

Therefore, the resistance change due to the faulty cell can be seen in the high-frequency response after processing the signals using a high-pass filter. Consequently, the resistance of the battery string can be estimated and used as the health indicator for fault diagnosis. Before designing the fault diagnosis algorithm, the validity of the simplified model for the high-frequency response of the battery string, which is obtained based on the first-order ECM, should be further verified through experiments or high-fidelity model. In this study, a physics-informed parameterized model for lithium-ion batteries, created by Yu et al. [20], is adopted as a virtual testbed. As shown in Fig. 3, this model is based on electrochemical theory, and every circuit element represents an actual physical phenomenon. The main elements include the resistance components associated with electrolyte ($R_{ele}$), current collector ($R_{cc}$), particle-to-current collector contact ($R_{p2cc}$), particle-to-particle contact ($R_{p2p}$), diffusion ($R_{diff}$), charge transfer ($R_{ct}$), solid electrolyte interphase (SEI) layer ($R_{sei}$), the capacitance components associated with double layer ($C_{dl}$), SEI layer ($C_{sei}$), and various particle-layer voltage sources ($E$). This model is scalable from a single particle model to a multi-particle model, and the particle layer number can also be customized. Due to space limitations, the model is not further described; detailed information about modeling and parameterization can be found in [20]. When compared to other high-fidelity models, the adopted cell model can be directly connected in parallel in Matlab/Simulink and therefore the battery string model can be established and studied.

To verify the relationship shown in Eq. (7), fresh and aged cells, which have significantly different parameters, are considered in one battery string. It is worth noticing that if Eq. (6) is verified, Eq. (7) can be deduced through mathematical induction. Four cells (5Ah lithium-polymer pouch cells [20]) with different degradation levels are used to form 6 different two-cell strings to verify Eq. (6). Typical parameters related to the degradation level of four selected cells are listed in Table I.

We point out that the fresh cell (i.e., Cell #1) is experimentally calibrated in [20] and the parameters of the most aged cell (i.e., Cell #4) are roughly set based on the estimation results in [20]. In addition, the parameters of Cell #2 and Cell #3 are artificially chosen for validation purposes, and it is assumed that all other parameters in the battery model do not change with degradation.

The current profiles used for resistance estimation are sinusoidal curves (with frequency of 0.5Hz and amplitude of 0.5C) combined with a DC component (with amplitude of 0.5C) to continuously discharge the battery from the initial SoC (i.e., 100%). A Butterworth high-pass filter with a 3dB bandwidth of 0.05Hz is adopted to extract the high-frequency dynamics, and a Kalman filter is used to estimate the resistance based on Eqs. (5) and (6).

First, the ohmic resistances of the single cells are characterized, as listed in Table I, and these values can be used to calculate the theoretical resistance of the battery strings based on Eq. (6). Furthermore, the sinusoidal current profiles are also used for the battery strings to estimate resistance, as listed in Table II. When comparing estimated and theoretical results, we find that the errors are small and therefore the relationship shown in Eq. (7) is proven to be accurate. Note that the resistance of the battery string specifically denotes the high-frequency resistance in this paper and it does not apply to the general response of the unfiltered signals $i_b$ and $v_b$. Detailed information about sequentially estimating battery parameters by separating its dynamics in the frequency domain is provided in [19]. In this paper, only the ohmic resistance is needed and therefore only the high-frequency response is used. The estimated ohmic resistance can converge to the actual value within 100s [19], and it is fast enough for the fault diagnosis. Note that the desired high-frequency signals can be injected in the charging process without perturbing the power supply.

## III. FAULT DIAGNOSIS FOR BATTERY STRINGS

### A. Cell Resistance Distribution

Based on the aforementioned analysis, the resistance of parallel battery strings can be used as the health indicator for the fault diagnosis. Given that the resistance increase of the battery string can be caused either by a cell fault or cell-to-cell variation, the resistance thresholds between normal and abnormal behavior in the fault diagnosis algorithm should be carefully designed to accurately detect the cell fault from normal cell-to-cell variations.

For fresh cells, variation is caused by manufacturing inconsistency, and this variation can enlarge as the battery cells age non-uniformly during operation. If the thresholds are set too tight, the normal variation may be detected as the fault, leading to a false alarm. On the other hand, if the thresholds are relaxed, the missed detection rate will be high. To balance the needs to avoid false alarms and missed detections, the optimal thresholds are necessary and can be obtained based on the statistical analysis. The offline design for the threshold selection and online implementation of the proposed fault diagnosis algorithm is shown in Fig. 4.

Schuster et al. [14] conducted extensive experiments on 484 fresh and 1908 aged lithium-ion cells, and characterized the capacity and impedance to quantify the cell-to-cell variation. It is verified that a normal distribution can be used for the cell parameters. Therefore, we assume the ohmic resistance of the individual battery cells is normal distribution $N(\mu, \sigma^2)$, where $\mu$ is the average resistance and $\sigma^2$ is the resistance variation. See Fig. 5 (a) and Fig. 6 (a) for a fresh and aged cell with $\mu$=6 mΩ and $\sigma$=0.12 mΩ for the new cell and $\mu$=11 mΩ and $\sigma$=0.385 mΩ for the aged one. It is found that the cell-to-cell variation, which can be quantified by the relative coefficient of variation $\kappa=\sigma/\mu$, increases with battery degradation [14]. Increasing variation inside the battery pack may be caused by uneven current distribution, which is directly caused by the resistance variation, and uneven temperature distributions, which may be caused by the resistance variation and the uneven cooling. The root causes for cell degradation and fault are beyond the scope of this paper.

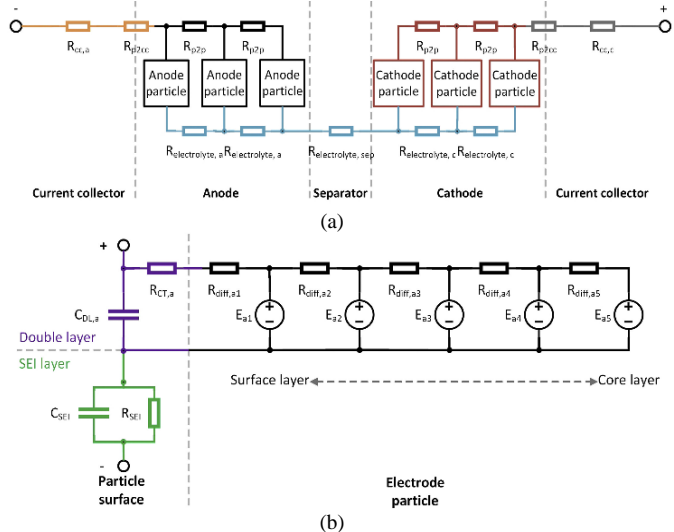

Figure 3. The physics-informed model for lithium-ion batteries. (a) Multi-particle model. (b) Anode particle subsystem. (With permission from Elsevier) [20]

TABLE I
Four cells with various degradation levels

| Cell number | Degradation in capacity (%) | $R_{sei}$ increase (%) | $R_{ele}$ increase (%) | $R_{ct}$ increase (%) | Ohmic resistance (mΩ) |
|---|---|---|---|---|---|
| Cell #1 | 0 | 0 | 0 | 0 | 5.8 |
| Cell #2 | 1.55 | 11 | 10 | 10 | 7 |
| Cell #3 | 3.25 | 23 | 20 | 10 | 7.2 |
| Cell #4 | 7.48 | 112 | 60 | 30 | 10.5 |

### B. Battery String Resistance Distribution

To investigate the resistance distributions of fresh and aged battery strings, we first focus on small battery strings with 5 cells. The resistance distributions of the battery strings are obtained through Monte Carlo simulations including 10000 samples. As shown in Fig. 5 (b), the resistance of the fresh battery strings can be approximately characterized by a normal distribution with a mean value $\mu$=1.2 mΩ and the coefficient of variation $\kappa$=0.89%, showing that the resistance distribution of battery strings becomes narrow when compared to that of individual fresh cells (where $\kappa$=2%).

Similarly, as shown in Fig. 6 (b), the mean value $\mu$=2.2 mΩ and the coefficient of variation $\kappa$=1.6% also decrease compared to aged cells, where $\kappa$=3.5%. Note that the selection of cell parameters is based on the experimental results provided in [14]. In addition, if the cell resistance is a normal distribution, theoretically the resistance of battery strings is not normal due to the nonlinear relationship shown in Eq. (7). However, a normal distribution is still assumed to simplify the analysis. Based on the distributions shown in Figs. 5 and 6, the thresholds for fault diagnosis can be chosen.

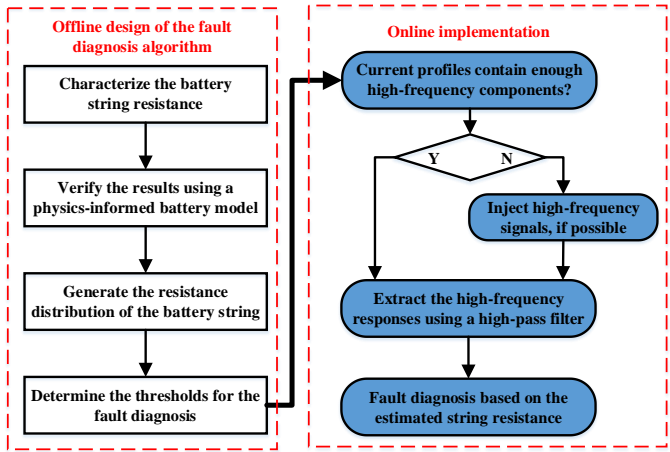

Figure 4. Schematic of the proposed fault diagnosis algorithm.

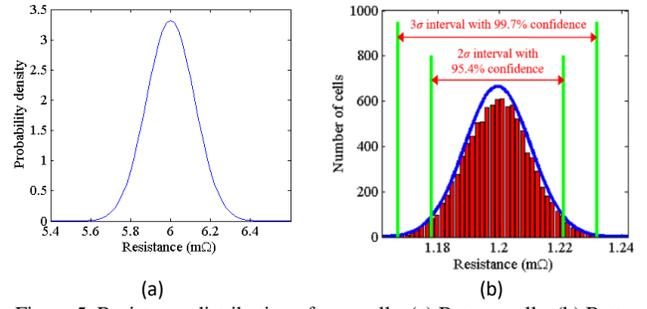

(a) (b)

Figure 5. Resistance distribution of new cells. (a) Battery cells. (b) Battery strings (5 cells).

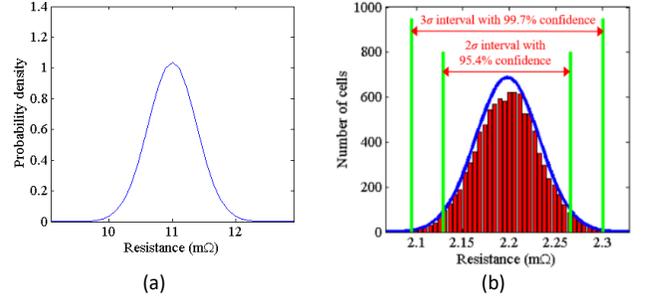

(a) (b)

Figure 6. Resistance distribution of aged cells. (a) Battery cells. (b) Battery strings (5 cells).

TABLE II
Validation results for the estimated resistance

| Battery strings | Estimated results (mΩ) | Theoretical results (mΩ) | Errors (%) |
| --- | --- | --- | --- |
| Cells #1 and #2 | 3.15 | 3.17 | 0.6 |
| Cells #1 and #3 | 3.2 | 3.21 | 0.3 |
| Cells #1 and #4 | 3.7 | 3.74 | 1.1 |
| Cells #2 and #3 | 3.55 | 3.55 | 0 |
| Cells #2 and #4 | 4.2 | 4.2 | 0 |
| Cells #3 and #4 | 4.25 | 4.27 | 0.5 |

For the studied fault (i.e., one faulty cell with a significant degradation), the upper threshold is for cell degradation related fault, and the lower threshold can be used to detect other faults such as the external/internal short circuits. As shown in Figs. 5 and 6, different intervals correspond to different confidence ratios and therefore directly determine the false alarm rates, which can be represented by the outliers (i.e., the red columns outside the green lines). If $2\sigma$ interval bounds are used as the thresholds, the possibility for a false alarm is 4.6%. Even though the larger interval (e.g., the $3\sigma$ interval) can reduce the false alarm rate, it will also increase the missed detection rate, as will be shown in the following.

### C. Performance of the Designed Fault Diagnosis Algorithm

In this subsection we evaluate the fault diagnosis performance for threshold of $2\sigma$. We first fix the battery string size (i.e., 5 cells) and evaluate the proposed algorithm for different fault levels (i.e., different resistance increases in the faulty cell). A fault means that there is at least a 60% increase in the cell resistance, as the battery fault in vehicle applications is defined to have at least 20% reduction in capacity or 60% increase in resistance [21]. Note that the proposed algorithm can be potentially used for health monitoring when the resistance increase of battery cell is less than 60%. As shown in Fig. 7, for fresh cells, all the faults can be detected because the resistances are all outside the 2σ region. The same simulation is conducted for aged battery strings, as shown in Fig. 8. Similarly, all faults can be detected, meaning that the missed detection rate is 0%. However, it can be seen that the resistance of the fault battery strings approaches the upper threshold when compared to the results of the fresh battery strings, as shown in Fig. 7. Note that the degradation related to the fault in this paper indicates the additional degradation of one specific cell when compared to the other cells, even though all cells may have the uniform degradation.

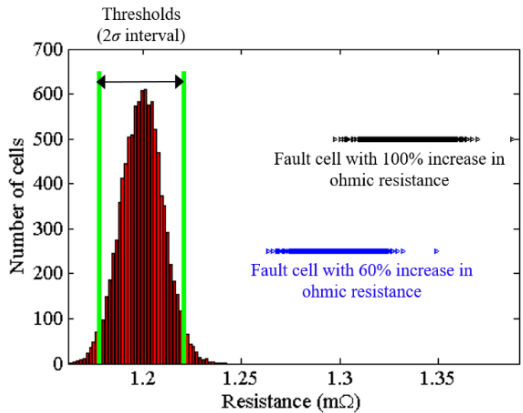

Figure 7. Resistance distribution of new battery strings (5 cells in parallel including one fault cell).

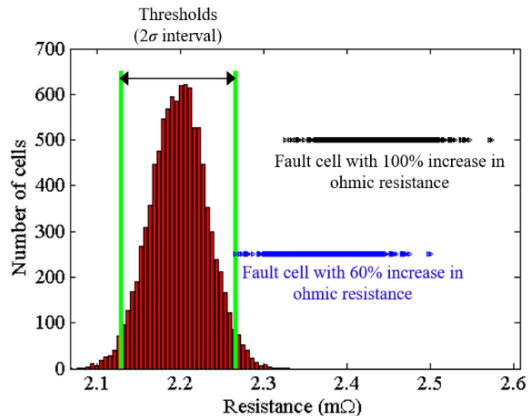

Figure 8. Resistance distribution of aged battery strings (5 cells in parallel including one faulty cell).

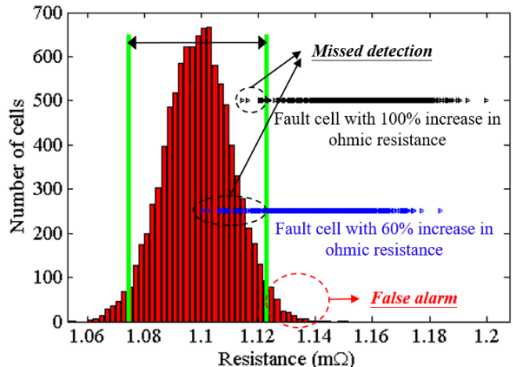

Figure 9. Resistance distribution of aged battery strings (10 cells in parallel including one faulty cell).

The size of the battery string (i.e., the cell number) also significantly influences the fault diagnosis performance. It is intuitive that, when the battery string is large, the fault is less obvious no matter what diagnosis algorithm is used. Focusing on aged battery cells, larger battery strings, including 10 cells, are investigated. As shown in Fig. 9, there are missed detections for both fault levels. When the faulty cell has the resistance increases of 60% and 100%, the missed detection rates are 7.25% and 0.4%, respectively. For example, for a battery string including 80 cells in total and one faulty cell with a 60% resistance increase, the missed detection rate is above 40%. When the battery string includes fewer than 20 cells, the resistance can be used as the health indicator and the fault diagnosis performance is satisfactory, given that the missed detection rate is low.

## IV. CONCLUSION

This paper proposes a fault diagnosis algorithm for parallel-connected battery cells, which includes one faulty cell with a significant degradation when compared to the other cells. The studied problem is challenging, since generally only one voltage sensor and one current sensor are used in one battery string. The lack of independent current sensors results in low detectability for the fault. The resistance of the battery string is selected as the health indicator, since the ohmic resistance of a battery cell significantly increases with degradation. Based on the high-frequency response of the battery string, a simplified fault detection-oriented model is derived and validated by a physics-informed battery model. The resistance distribution of battery strings is then characterized considering the resistance distribution of fresh and aged cells, which have different coefficients of variation. To balance false alarm and the missed detection rates, 2σ interval bounds are selected as the thresholds for fault diagnosis, resulting 4.6% false alarm rate. Simulation results show that the fault diagnosis for aged battery strings is more difficult given the larger coefficient of variation of the cell resistance distribution (e.g., aged cells). In addition, fault diagnosis for larger battery strings (i.e., more battery cells) is also difficult, as the resistance of the faulty cell will be compensated by other cells in the battery string.